\documentclass[aps,prl,twocolumn,10pt,longbibliography,superscriptaddress]{revtex4-2}
\usepackage{physics}
\usepackage{bbold}
\usepackage{multirow}
\usepackage[english]{babel}
\usepackage[utf8x]{inputenc}
\usepackage{amssymb}
\usepackage{amsmath}
\usepackage{dsfont}
\usepackage{comment}
\usepackage{lipsum}  
\usepackage[normalem]{ulem}
\usepackage[pdftex]{graphicx}
\usepackage{xspace}
\usepackage{dsfont}
\usepackage{bm}
\usepackage[export]{adjustbox}
\usepackage[pdfencoding=auto, psdextra]{hyperref}
\usepackage{cleveref}
\usepackage{longtable}
\usepackage[table]{xcolor}
\usepackage{float}
\hypersetup{
    colorlinks,%
    citecolor=blue,%
    filecolor=blue,%
    linkcolor=blue,%
    urlcolor=blue
}
\usepackage{tabularx}

\usepackage[usenames, dvipsnames]{xcolor}

\usepackage{soul}

\newcommand{\vteo}{V$_2$Te$_2$O\xspace}
\newcommand{\lacuo}{La$_2$CuO$_4$\xspace}

\newcommand{\stkout}[1]{\ifmmode\text{\sout{\ensuremath{#1}}}\else\sout{#1}\fi}

\begin{document}

\title{Persistent altermagnetism}

\author{Warlley H. Campos}
\email{Corresponding author: camposwa@pks.mpg.de}
\affiliation{Max Planck Institute for the Physics of Complex Systems, N\"othnitzer Str. 38, 01187 Dresden, Germany}
\author{F. C. Fobasso Mbognou}
\affiliation{Max Planck Institute for the Physics of Complex Systems, N\"othnitzer Str. 38, 01187 Dresden, Germany}
\author{Anna Birk Hellenes}
\affiliation{Institute of Physics, Czech Academy of Sciences, Cukrovarnická 10, 162 00 Praha 6, Czech Republic}
\author{Joseph Poata}
\affiliation{Max Planck Institute for the Physics of Complex Systems, N\"othnitzer Str. 38, 01187 Dresden, Germany}
\author{Taikang Chen}
\affiliation{Max Planck Institute for the Physics of Complex Systems, N\"othnitzer Str. 38, 01187 Dresden, Germany}
\author{Jan Priessnitz}
\affiliation{Max Planck Institute for the Physics of Complex Systems, N\"othnitzer Str. 38, 01187 Dresden, Germany}
\author{Libor Šmejkal}
\email{Corresponding author: lsmejkal@pks.mpg.de}
\affiliation{Max Planck Institute for the Physics of Complex Systems, N\"othnitzer Str. 38, 01187 Dresden, Germany}
\affiliation{Max Planck Institute for Chemical Physics of Solids, N\"othnitzer Str. 40, 01187 Dresden, Germany} 
\affiliation{Institute of Physics, Czech Academy of Sciences, Cukrovarnická 10, 162 00 Praha 6, Czech Republic}
\date{\today}

\begin{abstract}
Persistent spin textures with collinear spin polarization are promising platforms for spintronics applications. However, their typically relativistic spin–orbit origin leads to weak spin splittings and fragile spin coherence.
Here, we demonstrate a previously overlooked class of robust collinear spin polarization protected by mirror symmetry in combination with a strong exchange-driven altermagnetic order, which persists even in the presence of spin-orbit coupling. By combining first-principles calculations with a systematic classification of spin and magnetic layer groups, we identify this phenomenon—termed persistent altermagnetic spin polarization (PASP)—to occur in 158 spin layer groups and in representative materials including metallic V$_2$Te$_2$O, insulating La$_2$CuO$_4$, and semiconducting VSI$_2$. Furthermore, we theoretically demonstrate that PASP is ferroelectrically switchable in VSI$_2$. Finally, we show that this PASP switching can lead to large changes in spin-filtering conductance in a model all-altermagnetic junction.
Our results open the possibility of employing PASP in all-altermagnetic magnetic memory and spin-transistor devices and establish universal principles of altermagnetism in spin–orbit-coupled monolayers.
\end{abstract}

\maketitle

Relativistic spin-orbit coupling (SOC) in noncentrosymmetric systems typically generates noncollinear spin textures in crystal momentum space in the form of Rashba~\cite{bychkov1984oscillatory}, Dresselhaus~\cite{dresselhaus1955spin}, or Weyl textures~\cite{zollner2025first}. 
Thus, SOC generally disrupts spin collinearity and shortens spin lifetimes~\cite{mi1986spin}, limiting spintronic applications such as spin field-effect transistors~\cite{datta1990electronic}. To overcome these limitations, an early strategy proposed balancing the Rashba and Dresselhaus parameters in quantum wells to produce a momentum-independent collinear relativistic spin polarization~\cite{schliemann2003nonballistic}. This configuration was predicted to support a helical spin density wave with an extended spin lifetime~\cite{bernevig2006exact}, dubbed persistent spin helix, which was later confirmed experimentally via transient spin-grating spectroscopy~\cite{koralek2009emergence} and magneto-optical Kerr rotation~\cite{walser2012direct}. However, achieving such precise tuning of relativistic effects is experimentally demanding, as it requires accurate control over the quantum well width, doping concentration and external electric field~\cite{tao2018persistent}.

\begin{figure}[h!]
    \centering
    \includegraphics[width=\linewidth]{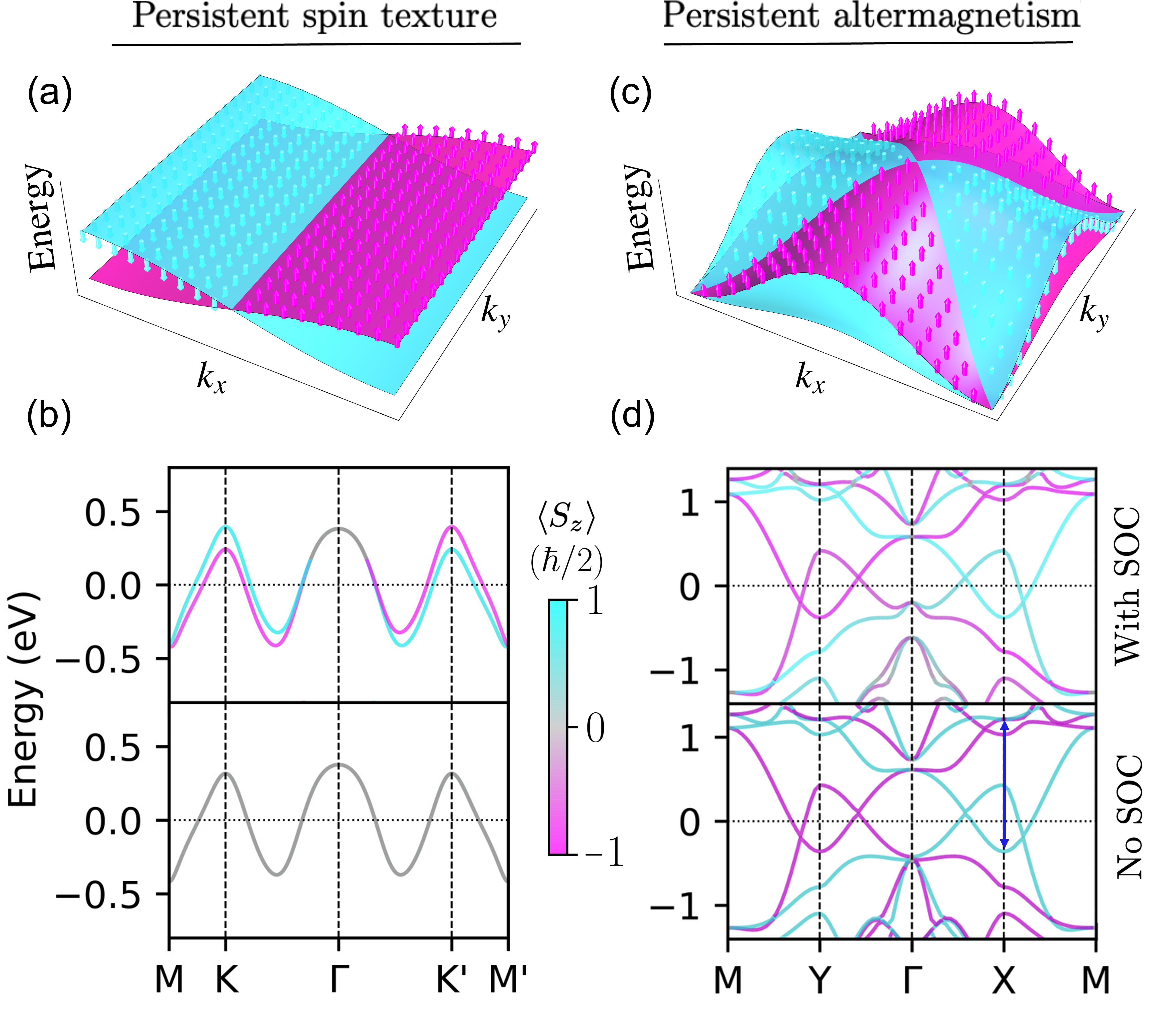}
    \caption{Persistent spin texture versus persistent altermagnetic spin polarization (PASP). (a) Schematic illustration of the persistent spin texture induced by weak Ising spin-orbit coupling (SOC). (b) Electronic band structures of NbSe$_2$ with (top) and without (bottom) SOC. %(c) Crystal structure of NbSe$_2$, with Nb (Se) atoms represented in gray (blue). 
    (c) PASP in a $d$-wave Lieb-lattice altermagnet (AM). (d) Band structure of the Lieb-lattice AM V$_2$Te$_2$O with (top) and without (bottom) SOC. The large altermagnetic splitting of $\sim1.5$~eV is indicated by the blue double arrow in (d).  The positive (negative) spin expectation value $\langle S_z \rangle$ is highlighted in magenta (cyan) color.} 
    \label{fig:fig0}
\end{figure}%

\begin{figure*}[t]
    \centering
    \includegraphics[width=\linewidth]{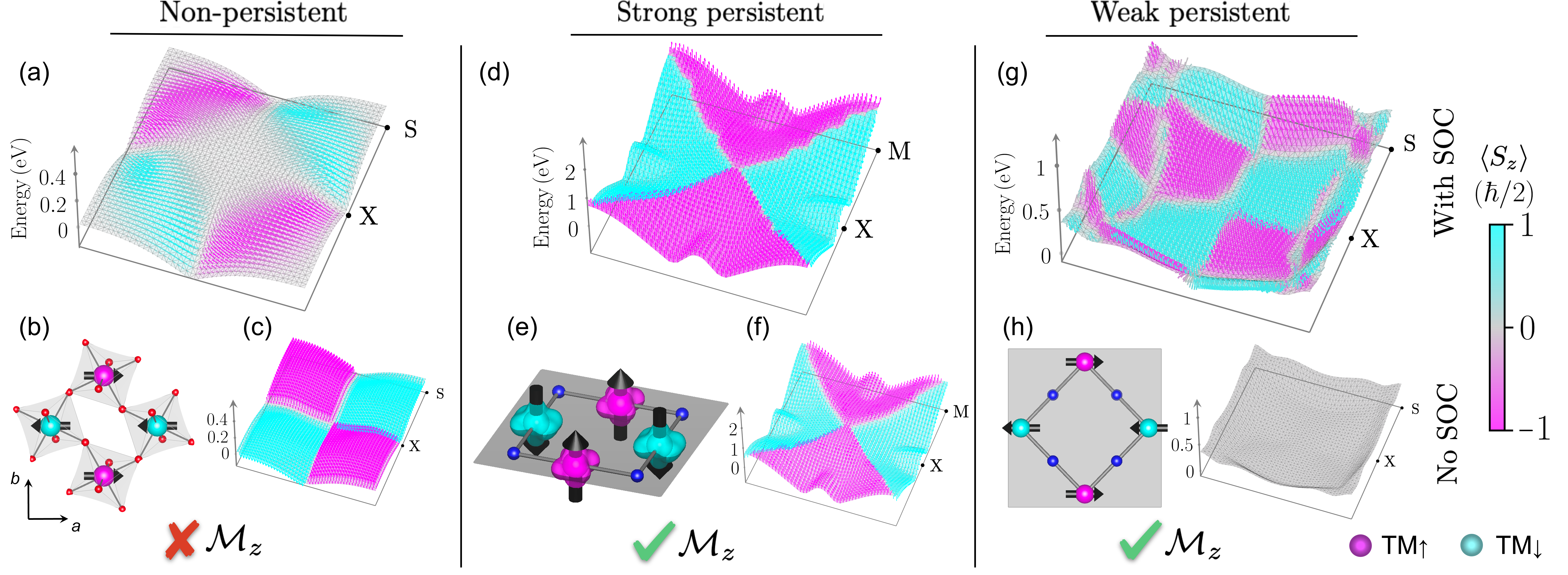}
    \caption{Three types of  spin polarization in AMs with SOC. (a) Non-collinear relativistic spin texture of the topmost valence band of the non-persistent AM OsF$_4$. (b) Magnetic structure of OsF$_4$. (c) Nonrelativistic spin expectation value for the same band.
    (d) Strong PASP in the selected band at the Fermi level of the strong persistent AM \vteo. (e) Magnetic structure of \vteo. (f) Nonrelativistic spin expectation value with large altermagnetic splitting. (g) Weak PASP in the topmost valence band of the weak persistent AM \lacuo. (h) Magnetic structure of \lacuo. (i) Nonrelativistic band dispersion with spin degeneracy enforced by the $[C_2||M_z]$ symmetry. The gray plane $\mathcal{M}_z$ in (e,h) denotes the horizontal mirror plane. 
    Transition-metal atoms ($\mathrm{TM} = \mathrm{Os},\mathrm{V},\mathrm{Cu}$) with opposite spin orientations are shown in magenta (TM$\uparrow$) and cyan (TM$\downarrow$). The Te and La atoms in (e) and (h) are omitted for clarity. See~\cite{SM} for the full crystal structures.}
    \label{fig:Fig2}
\end{figure*}%

Later studies have shown that collinear spin polarization can be enforced around certain high-symmetry points in the Brillouin zone (BZ) by nonsymmorphic~\cite{tao2018persistent} or mirror~\cite{ji2022symmetry,campos2024dual} symmetries. This mechanism, so-called persistent spin texture, has recently undergone symmetry classification for both three-dimensional (3D)~\cite{tao2018persistent,tan2025understanding,kilic2024universal} and two-dimensional (2D)~\cite{ji2022symmetry} materials. 
In transition-metal dichalcogenides (TMDs)~\cite{bawden2016spin}, Ising SOC can lead to a time-reversal-symmetric persistent spin texture, as illustrated in Fig.~\ref{fig:fig0}(a) for energy bands in momentum space. Although persistent spin textures represent an important step toward achieving intrinsically long spin lifetimes, they are typically limited by very weak spin splitting, since they arise from relativistic corrections to the electronic structure. For example, Fig.~\ref{fig:fig0}(b) shows that the relativistic spin splitting in the TMD NbSe$_2$~\cite{bawden2016spin} is in the order of $\sim 0.01-0.1$~eV. 

Here, we propose an alternative strategy that circumvents the limitations  imposed by weak spin splitting and quantum well engineering. Our approach consists of realizing mirror-symmetry-protected collinear spin polarization in pristine 2D altermagnets (AMs) with intrinsically large spin splitting. AMs are characterised by alternating spin polarization constrained by spin symmetries in both real and momentum spaces~\cite{vsmejkal2022beyond,vsmejkal2022emerging}. Although large altermagnetic splittings in the order of $\sim 0.1-1$~eV~\cite{vsmejkal2022beyond,vsmejkal2022emerging,jaeschke2025atomic,parthenios2025spin} have recently been experimentally confirmed in materials such as MnTe~\cite{krempasky2024altermagnetic,lee2024broken,amin2024nanoscale} and CrSb~\cite{reimers2024direct},  SOC typically generates noncollinear spin textures
~\cite{milivojevic2024interplay}.

By performing a systematic classification using magnetic and spin layer groups (MLGs and SLGs)~\cite{litvin2013magnetic}, we identified a family of 2D AMs with altermagnetic splitting and robust collinear spin polarization protected by mirror symmetry in the presence of SOC. We refer to this phenomenon as \emph{persistent altermagnetic spin polarization} (PASP) and the corresponding magnetic state as \emph{persistent altermagnetism}.
In Fig.~\ref{fig:fig0}(c), we illustrate the PASP in a $d$-wave Lieb-lattice AM. As shown by the electronic band structure in Fig.~\ref{fig:fig0}(d), the large AM splitting in the order of $\sim 1.5$~eV is of nonrelativistic exchange origin (lower panel), and its out-of-plane collinearity is preserved and protected by mirror symmetry once  SOC is taken into account (top panel).

We classify all 2D persistent AMs into two types
according to the character of the associated spin splitting. A strong persistent AM is characterized by a nonrelativistic altermagnetic splitting of exchange origin, while a \textit{weak persistent AM} exhibits SOC-assisted spin splitting which emerges on top of a nonrelativistic nodal plane. The terminology of strong versus weak spin splitting types was introduced in Ref.~\cite{krempasky2024altermagnetic} for bulk 3D MnTe.

\emph{Strong and weak persistent altermagnetism---} Before presenting material candidates for persistent altermagnetism, we first illustrate how SOC can compromise spin collinearity in AMs without PASP. Fig.~\ref{fig:Fig2}(a) shows the highly noncollinear relativistic spin texture of the topmost valence band of OsF$_4$~\cite{sodequist2024two}, providing an example of a non-persistent AM. This behavior arises from the absence of horizontal mirror symmetry in the magnetic group of OsF$_4$ [Fig.~\ref{fig:Fig2}(b)]. Although the corresponding nonrelativistic valence band [Fig.~\ref{fig:Fig2}(c)] exhibits exchange-driven collinear altermagnetic splitting, the inclusion of SOC distorts this alignment in the absence of symmetry protection, producing the intricate relativistic spin texture in Fig.~\ref{fig:Fig2}(a).

In strong persistent AMs, the large AM splitting originates from nonrelativistic exchange coupling, and its collinearity is preserved in the presence of SOC due to the protection by mirror symmetry in the magnetic group. Relying on our symmetry classification, we identify the quasi-2D AM \vteo~\cite{li2024strain,cui2023giant} as a material candidate for the observation of strong PASP. Its 3D counterpart, previously investigated both theoretically~\cite{skornyakov2024coulomb} and experimentally~\cite{ablimit2018v2te2o}, has a layered van der Waals (vdW) structure, which facilitates the production of the quasi-2D altermagnetic phase via exfoliation.

In Fig.~\ref{fig:Fig2}(d), we show the momentum-resolved spin expectation value of \vteo in the presence of SOC for a selected band at the Fermi level, obtained from our density functional theory (DFT) calculations [see Supplementary Material (SM)~\cite{SM}] using the Vienna Ab-Initio Simulation Package (VASP)~\cite{Kresse1996Jul}. The corresponding magnetic structure in real space is presented in Fig.~\ref{fig:Fig2}(e). The spin expectation value without SOC for the same energy band is shown in Fig.~\ref{fig:Fig2}(f), revealing that the altermagnetic splitting originates from nonrelativistic exchange coupling. The spin point group (SPG) symmetries $[C_2||M_{110}]$ and $[C_2||M_{\bar{1}10}]$ (where operations on the left (right) of the double vertical bar act on the spin (real) space~\cite{vsmejkal2022beyond}) enforce the nodal lines along crystal momentum $k_y=\pm k_x$ in Fig.~\ref{fig:Fig2}(f). In Fig.~\ref{fig:Fig2}(d), the collinear altermagnetic spin polarization remains strictly preserved when SOC is included. This robustness originates from the mirror symmetry $M_z$ from the magnetic group, which enforces $\langle S_{x(y)} \rangle=0$ throughout the entire 2D BZ~\cite{campos2024dual}. The magnetic point group (MPG) symmetries $M_{\bar{1}10}$ and $M_{110}$ are responsible for enforcing the nodal lines along $k_y=\pm k_x$ in the presence of SOC.

Although weak persistent AMs also exhibit mirror-symmetry-protected collinear PASP, their nonrelativistic bands are completely spin-degenerate. This degeneracy, imposed by SPG symmetries, is lifted when SOC is included, leading to a typically weak relativistic spin splitting. From our symmetry classification, we identified the quasi-2D AM cuprate \lacuo as a realistic weak persistent AM candidate. In its bulk form, \lacuo was studied extensively as the parent compound of unconventional superconductivity~\cite{lane2018antiferromagnetic,pickett1989electronic,yu1987electronically,mattheiss1987electronic}, and its quasi-2D counterpart can be potentially exfoliated from the original 3D layered compound~\cite{lane2018antiferromagnetic}.

%\begin{widetext}

%\end{widetext}
\begin{figure}
    \centering
    \includegraphics[width=\linewidth]{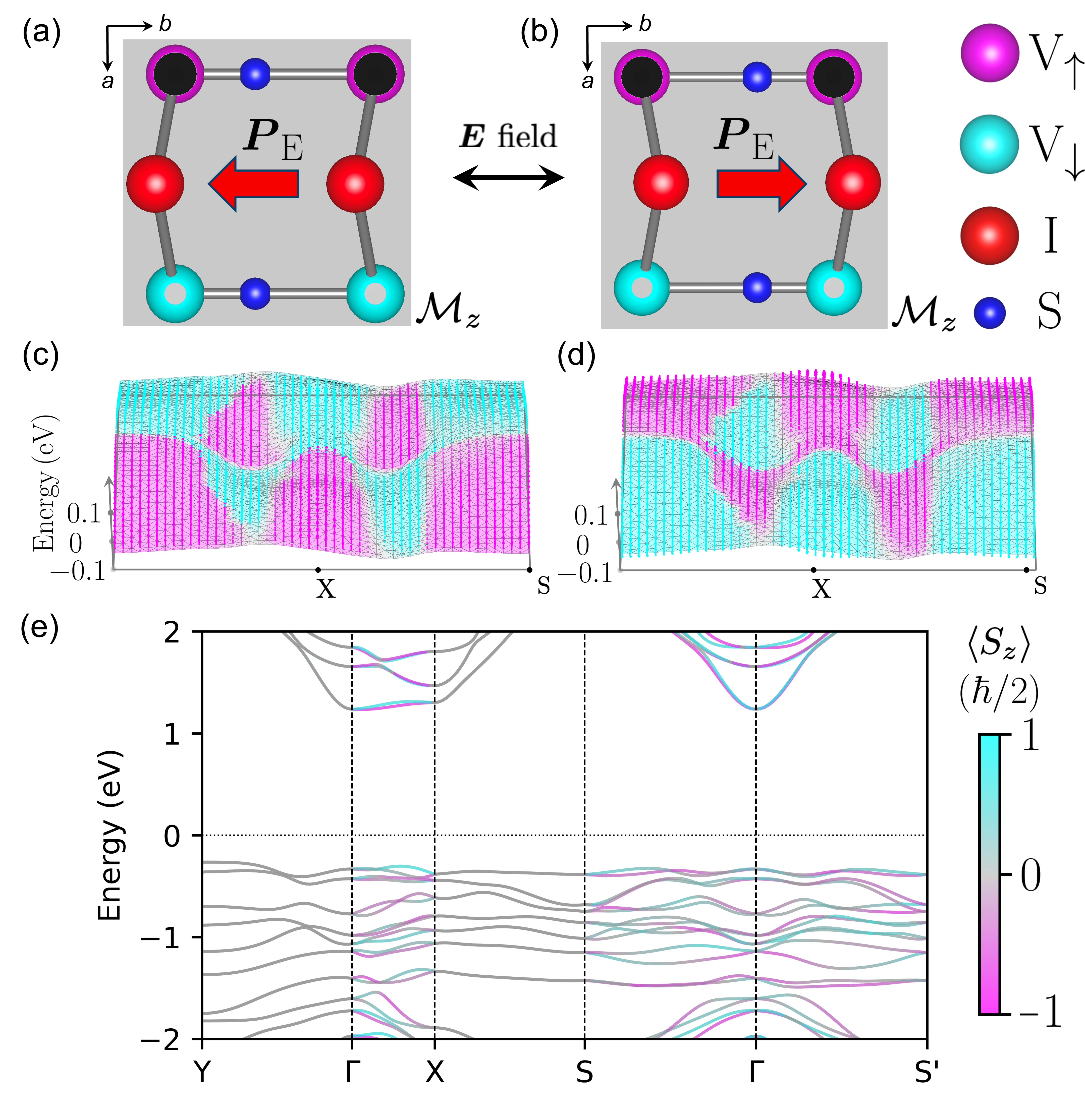}
    \caption{Altermagnetoelectric effect with strong PASP in the ferroelectric AM VSI$_2$. (a,b) Crystal and magnetic structures of VSI$_2$ for opposite directions of the ferroelectric polarization $\boldsymbol{P}_{\rm{E}}$, switchable by an in-plane electric field (black double arrow). V atoms with positive (negative) out-of-plane magnetic moments are shown in magenta (cyan). O (S) atoms are shown in blue (red). Red arrows indicate the direction of $\boldsymbol{P}_{\rm{E}}$. (c) and (d) show the strong PASP of the topmost valence band for the configurations in (a) and (b), respectively. (e) Relativistic  band structure corresponding to the configuration in (b,d).}
    \label{fig:fig3}
\end{figure}

Our DFT result for the spin expectation value of the relativistic topmost valence band of \lacuo is shown in Fig.~\ref{fig:Fig2}(g). The corresponding magnetic structure with N\'{e}el vector $\bm{N}\parallel[100]$ is presented in Fig.~\ref{fig:Fig2}(h). Here, the point group mirror symmetry $M_z$ protecting the PASP is associated with a space group glide mirror symmetry $M_z\bm{t}$, where $\bm{t}=(\frac{1}{2},\frac{1}{2},0)$ is a half-lattice fractional translation. 
The $d$-wave spin polarization is consistent with the mirror symmetries $M_x$ and $M_y$,  which enforce the nodal lines along $k_x=0$ and $k_y=0$ directions, respectively. Without SOC, \lacuo is invariant under the SPG symmetry $[C_2||M_z]$, which enforces spin degeneracy of the entire 2D BZ. Consistent with this symmetry constraint, our DFT calculations reveal an entirely spin-degenerate electronic structure in the absence of SOC [Fig.~\ref{fig:Fig2}(i)], confirming the presence of weak PASP in \lacuo.

\emph{Altermagnetoelectric effect with PASP---}We now demonstrate that PASP can arise in ferroelectric AMs~\cite{vsmejkal2024altermagnetic,gu2025ferroelectric} hosting the recently proposed altermagnetoelectric (AME) effect~\cite{vsmejkal2024altermagnetic}. In these systems, switching the ferroelectric polarization $\boldsymbol{P}_{\rm{E}}$ reverses the spin polarization while preserving the orientation of the Néel vector. Our symmetry classification identifies the 2D ferroelectric AMs VOX$_2$ and VSX$_2$ (X = Cl, Br, I, S)~\cite{zhu2025two} as promising candidates for strong PASP. Figs.~\ref{fig:fig3}(a,b) show the crystal and magnetic structures of VSI$_2$ for the two opposite ferroelectric polarizations, which can be switched by an in-plane electric field. The corresponding PASP of the topmost valence band is presented in Figs.~\ref{fig:fig3}(c,d), revealing a direct reversal of the spin polarization upon ferroelectric polarization switching. The relativistic band structure in Fig.~\ref{fig:fig3}(e) further confirms the PASP of the switched configuration. These results establish VSI$_2$ as a strong persistent AM in which the sign of the spin polarization can be switched electrically.

\begin{figure}[t]
    \centering
    \includegraphics[width=\linewidth]{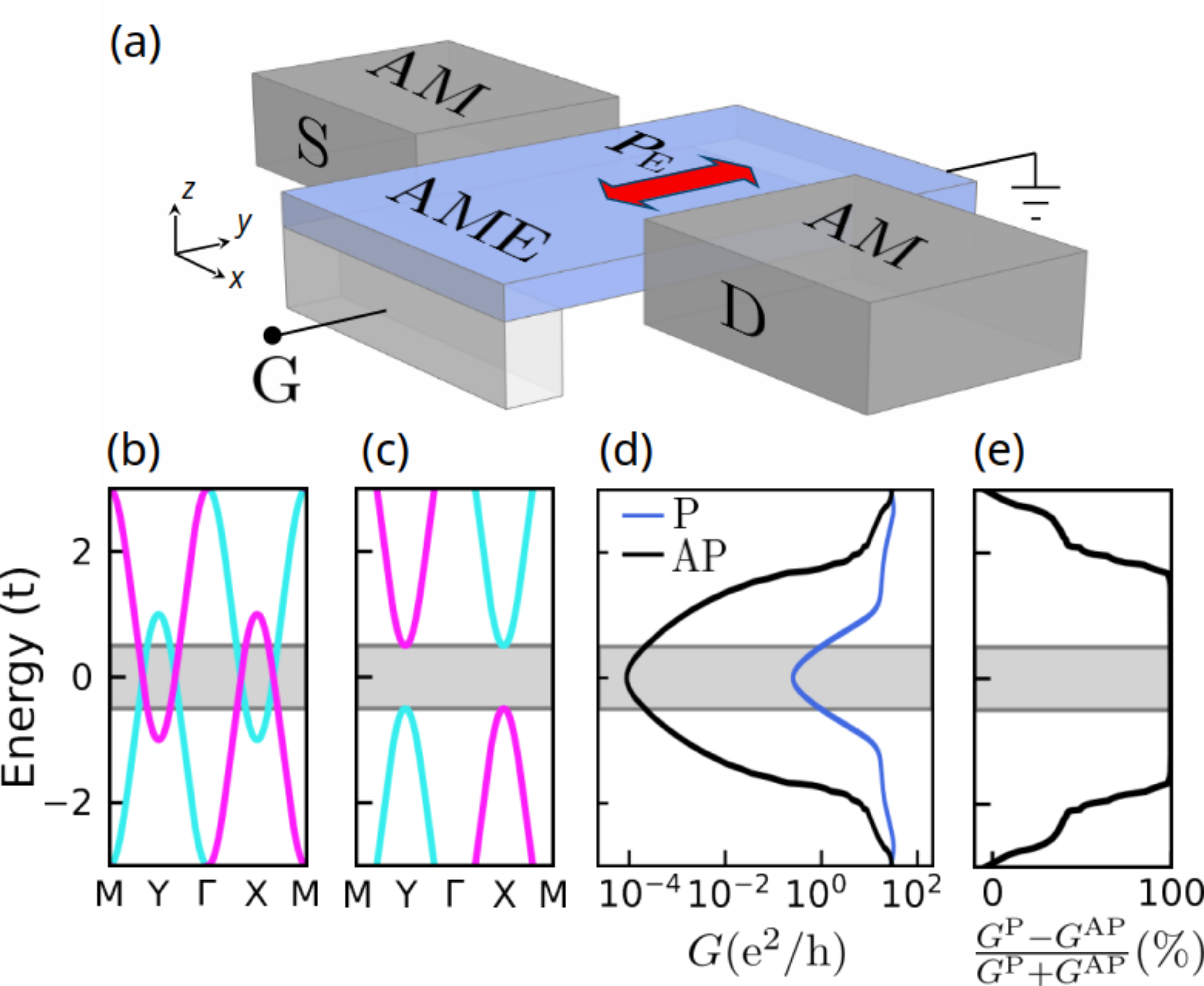}
    \caption{
     Spin-filtering junction based on electrically switchable PASP. (a) Schematic of device: the source (S) and drain (D) leads (gray) are composed of identical metallic AMs with fixed PASP aligned along the $\hat{z}$ direction. An altermagnetoelectric (AME) hosting strong PASP forms the central scattering region (blue). (b) and (c) show the energy bands of the  model describing the 2D bulk metallic AM leads and the semiconducting AME with switchable ferroelectric polarization $\boldsymbol{P}_{\rm{E}}$, respectively. (d) Calculated zero-temperature linear conductance $G$ for parallel (P, blue line) and antiparallel (AP, black line) mutual orientation of the leads and AME region.  (e) Computed tunneling magnetoresistance ratio.}
    \label{fig:transistor}
\end{figure}

\emph{Altermagnetic spin-filtering junction---} We now consider a tunnel junction memory device in which the magnetic state is written via the AME effect and read out through tunneling magnetoresistance. 
Fig.~\ref{fig:transistor}(a) illustrates the device geometry. The source (S) and drain (D) consist of identical metallic strong persistent AMs with fixed PASP. The metallic AM $d$-wave lead can be used as spin polarizer, as $d$-wave AMs exhibit strong nonrelativistic longitudinal spin polarized currents~\cite{vsmejkal2022giant,gonzalez2021efficient}. 
The central region of the device is based on the ferroelectric AM with the AME effect~\cite{vsmejkal2024altermagnetic,zhu2025two}. 
A lateral electrostatic gate (G)~\cite{gan2013high,xi2016gate} enables switching the in-plane ferroelectric polarization, $\boldsymbol{P}_{\rm{E}}$, between opposite orientations, thereby reversing the collinear PASP in the scattering region.

To model the tunneling magnetoresistance of our junction, we employ a minimal four-band tight-binding Hamiltonian (see Refs.~\cite{vsmejkal2022beyond,vsmejkal2022giant} and SM~\cite{SM}) that, by using different sets of parameters, captures both the metallic AM leads and the semiconducting AM hosting the AME effect. Figs.~\ref{fig:transistor}(b) and \ref{fig:transistor}(c) show the band structures describing the 2D bulk metallic leads and the semiconducting valleys~\cite{reichlova2024observation,ma2021multifunctional} of the AME-active region, respectively.

The zero-temperature linear-response conductance $G$ of our junction was calculated using the Kwant package~\cite{groth2014kwant} and is shown in Fig.~\ref{fig:transistor}(d). The insulating AME serves as a spin-filtering tunneling barrier~\cite{samanta2025spin}. The PASP of the leads is fixed and points along the $\hat{z}$ direction. When $\boldsymbol{P}_{\rm{E}}||\hat{y}$, the spin polarization in the scattering region aligns parallel to the PASP of the leads, resulting in a high conductance. Reversing the polarization to $\boldsymbol{P}_{\rm{E}}||-\hat{y}$ leads to an antiparallel orientation between the spin polarization in the central region and that of the leads and strongly suppresses the conductance. The resulting spin-filtering behavior~\cite{samanta2025spin} is summarized in Fig.~\ref{fig:transistor}(e), where the tunneling magnetoresistance ratio approaches  $\sim 100\%$ over the energy range $(-1.5t,\,1.5t)$, where $t$ is the strength of the spin-dependent hopping term in the model Hamiltonian. This electrically controlled reversal of PASP therefore enables a purely altermagnetic spin-filtering functionality.

\begin{table}[t]
    \centering
    \renewcommand{\arraystretch}{1.2}
    \begin{tabular}{|c|c|c|c|}
        \hline
        \textbf{Type} & \multicolumn{3}{c|}{\textbf{Subclass}} \\
        \hline
        \begin{tabular}{c}
             Strong\\
             PASP\\
            (92)
        \end{tabular}
        &  
        \begin{tabular}{c} 
            $d$-wave (69) \\ 
            \includegraphics[width=1cm]{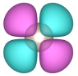} \\ 
            $\mathrm{V_2Te_2O}$ 
        \end{tabular}
        &
        \begin{tabular}{c} 
            $g$-wave (12) \\ 
            \includegraphics[width=1cm]{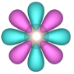} \\ 
            $\mathrm{FeS_2}$~\cite{li_altermagnetism_2025} 
        \end{tabular}
        & 
        \begin{tabular}{c} 
            $i$-wave (11) \\ 
            \includegraphics[width=1cm]{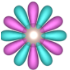} \\ 
            $\mathrm{MnP(S,Se)_3}$~\cite{mazin2023inducedmonolayeraltermagnetismmnpsse3} 
        \end{tabular}
        \\
        \hline

        \multirow{3}{*}{
        \begin{tabular}{c}
         Weak \\
         PASP \\
         (66)   
        \end{tabular}
        }
        &
        \multicolumn{3}{c|}{$[C_2||C_{2z}]$ (16)} \\
        \cline{2-4}
        &
        \multicolumn{3}{c|}{$[C_2||M_z]$ (27)} \\
        \cline{2-4}    
        &
        \multicolumn{3}{c|}{$[C_2||C_{2z}]$ \& $[C_2||M_z]$ (23)} \\
        \hline
        \end{tabular}
    \caption{Spin group classification of persistent AMs into strong and weak types. The strong type is further categorized into subclasses of different partial-wave character~\cite{jaeschke2025atomic}. The weak type is categorized into subclasses according to the symmetry protecting the nonrelativistic spin degeneracy in the 2D BZ. In parentheses, we show the total number of spin layer groups belonging to each type and subclass. The complete enumeration is provided in the SM~\cite{SM}.}
    \label{tab:table1_transposed}
\end{table}

\textit{Relativistic magnetic symmetry criteria for PASP---}
Achieving relativistic collinear spin polarization requires spin-split energy bands, except along nodal lines, and the spin expectation value globally constrained along the $z$ direction. These requirements are satisfied if the MPG fulfills two conditions~\cite{ji2022symmetry}: (1) it does not include $\mathcal{PT}$ symmetry and (2) it includes mirror symmetry $M_z$ along the out-of-plane $z$ direction.

Condition (1) prevents Kramers spin degeneracy and is automatically satisfied by AMs since $\mathcal{PT}$ symmetry is excluded by construction in these crystals~\cite{vsmejkal2022beyond}.  Condition (2) enforces spin alignment along the $z$ axis throughout the entire 2D BZ. Note that $M_z\bm{k}=\bm{k}$ in 2D, where $\bm{k}=(k_x,k_y)$, while $M_{z}(\langle S_x \rangle, \langle S_y \rangle, \langle S_z \rangle)=(-\langle S_x\rangle, -\langle S_y \rangle, \langle S_z\rangle)$. Therefore, invariance under $M_z$ imposes $\langle S_{x(y)}\rangle=0$  at every $\bm{k}$, and only the out-of-plane spin component $\langle S_z \rangle$ can be finite~\cite{campos2024dual}. In summary, in the presence of SOC, any 2D AM with either mirror or glide-mirror symmetry along the out-of-plane direction hosts PASP.

In the SM~\cite{SM}, we list all MPGs and corresponding MLGs~\cite{litvin2013magnetic} compatible with PASP. 
Furthermore, these groups can be polar, enabling the coexistence of PASP with ferroelectricity and the magnetoelectric effect~\cite{vsmejkal2024altermagnetic}.

\textit{Spin layer group classification of 2D AMs---}
To distinguish between the strong and weak PASP types, we introduce spin layer groups (SLGs), in which we extend the classification of the different nonrelativistic spin splittings in 2D AMs~\cite{zeng2024description} and antiferromagnetic surfaces~\cite{lange2026emergent}. SLGs combine 3D SPG operations~\cite{litvin1974spin} with 2D Bravais lattices. We construct all collinear SLGs by combining nonmagnetic layer group operations in the real space with the identity $E$ and a two-fold spin space rotation $C_2$~\cite{vsmejkal2022beyond} (see the SM~\cite{SM} for a detailed discussion of SLGs). Next, we identify and enumerate a total of 158 altermagnetic SLGs, which constitute the type-III AM spin group, ${\textbf{{G}}^{III} = [E|| \textbf{H}] + [C_{2}||A\textbf{H}]}$, in analogy to the classification of 3D AMs~\cite{vsmejkal2022beyond}. \textbf{H} is the halving subgroup of the corresponding spin group \textbf{G}, which contains the real space identity operation, and $A\textbf{H}=\textbf{G}-\textbf{H}$ contains all remaining operations of \textbf{G}, which are not in \textbf{H}. $A$ is a crystallographic spin transposing operation. The full enumeration and classification including the type-I ferromagnetic and type-II antiferromagnetic spin layer groups is provided in the SM~\cite{SM}.

Furthermore, we divide the type-III AM spin groups into two classes of SLGs according to the nodal structure of the energy bands. The first class is that of  strong AMs, which contains 92 SLGs and is characterized by the absence of a nodal plane coinciding with the plane of the 2D BZ. We further subdivide these 92 SLGs into $d$-, $g$-, or $i$-wave subclasses, according to the number of nodal planes perpendicular to the 2D BZ, \emph{i.e.}, two, four or six nodal planes, respectively.
In Table~\ref{tab:table1_transposed}, we provide the number of SLGs and realistic material candidates for each subclass.

The second class, of weak AMs, is characterized by a nodal plane coinciding with the 2D BZ and contains 66 SLGs. This degeneracy is enforced by spin symmetry operations that flip spin while preserving the 2D crystal momentum. In Table~\ref{tab:table1_transposed}, we further classify weak AMs into three subclasses according to which symmetry enforces the degeneracy: $[C_2||C_{2z}]$,  $[C_2||M_z]$ or both. We also provide the number of SLGs for each subclass.

\emph{Discussion---}
Our classification using MLGs and SLGs provide a unified understanding of persistent altermagnetism in 2D. For a given crystal with collinear compensated magnetic order, its MLG determines if it hosts PASP, while its SLG indicates if the PASP is of strong or weak type. This combined analysis allows for the identification of a 2D AM as a non-persistent AM or a persistent AM of strong or weak type. In the SM~\cite{SM}, we compile a list of 2D altermagnetic candidates~\cite{bai2024altermagnetism} in which PASP is allowed by symmetry.

We point out that the spin-degenerate nodal structure in the $k_z=0$ plane can differ from the global 3D partial wave character. For instance, AMs such as MnP(Se,S)$_3$~\cite{mazin2023inducedmonolayeraltermagnetismmnpsse3} or Fe$_2$O$_3$~\cite{hoyer2025altermagnetic} exhibit bulk $g$-wave spin Laue group, but the $k_z=0$ plane hosts an extra nonrelativistic crossing, leading to $i$-wave planar nodal structure. Furthermore, we note that weak PASP is analogous to the states at the $k_z=0$ plane of the 3D BZ of MnTe~\cite{krempasky2024altermagnetic} and monolayer Hf$_2$S~\cite{bai2025anomalous}. While in MnTe the resulting state is not a pure altermagnetic $d$-wave, its collinear spin polarization is persistent and protected by the in-plane mirror plane and was experimentally observed by spin-polarized angular photoemission spectroscopy~\cite{krempasky2024altermagnetic,din2025unconventional}.

The electrically switchable PASP can be exploited to encode a nonvolatile memory bit~\cite{vsmejkal2022giant,shao2024antiferromagnetic}. Alternatively, the device proposed in Fig.~\ref{fig:transistor} can operate as a spin transistor, as the tunneling conductance in the gap region of the AME-active material is approximately $10^{4}$ times larger in the parallel (``on-state'') than in the antiparallel (``off-state'') configuration.

Conventional spin field-effect transistors are often limited by weak SOC. This concept was recently extended to employ an AME material sandwiched between ferromagnetic electrodes with an out-of-plane gating geometry~\cite{zhu2025altermagnetoelectric}. In contrast, our proposed spin-transistor junction is constructed entirely from altermagnetic terminals and operated via an in-plane gating geometry. For metallic leads, Lieb lattice AMs of the \vteo family can be employed, while the ferroelectric region can be realized using materials from the VSI$_2$ family.

\emph{Conclusions---}
In this manuscript, we proposed the realization of collinear PASP protected by mirror symmetry in a class of 2D altermagnets which we term \emph{persistent altermagnets}. Using state-of-the-art spin and magnetic layer group analysis, we classified all magnetic layer groups and showed that persistent altermagnetism can be achieved in 76 out of 528 relativistic magnetic layer groups built from 92 (strong type) and 66 (weak type) spin layer groups belonging to the altermagnetic type. Density functional theory calculations for the strong and weak persistent altermagnets \vteo and \lacuo, respectively, corroborate our symmetry-based predictions. We have also demonstrated PASP in altermagnetoelectric VSI$_2$.  Finally, we have proposed an all-altermagnetic spin-filtering junction, where PASP can be controlled by electric switching and read-out by a large spin-filtering tunneling magnetoresistance. Such a device offers a promising platform for the realization of altermagnetic memory and transistor applications.

\section{Acknowledgments}
The authors acknowledge support from the ERC Starting Grant No. 101165122, the ERC Advanced Grant no. 101095925, and funding from the Deutsche Forschungsgemeinschaft (DFG) grant no. TRR 173 268565370 (project A03).

%\bibliography{library}
%apsrev4-2.bst 2019-01-14 (MD) hand-edited version of apsrev4-1.bst
%Control: key (0)
%Control: author (8) initials jnrlst
%Control: editor formatted (1) identically to author
%Control: production of article title (0) allowed
%Control: page (0) single
%Control: year (1) truncated
%Control: production of eprint (0) enabled
%

\end{document}